\documentclass[letter]{aa}
\pdfoutput=1
\usepackage{graphicx}
\usepackage{txfonts}
\usepackage{multirow}
\usepackage{natbib}
\usepackage{color}
\usepackage{amsmath}
\begin{document}

   \title{Successful application of PSF-R techniques to the case of the globular cluster NGC~6121 (M~4)\thanks{Based on observations obtained under the program ID 60.A-9801(S)}}

   \subtitle{}

   \author{D. Massari\inst{1,2,3}
          \and
          A. Marasco\inst{4,3}
          \and
          O. Beltramo-Martin\inst{5,6}
          \and
          J. Milli\inst{7,8}
          \and
          G. Fiorentino\inst{9}
          \and
          E. Tolstoy\inst{3}
          \and
          F. Kerber\inst{10}         
          }

    \institute{
             Dipartimento di Fisica e Astronomia, Universit\`{a} degli Studi di Bologna, Via Gobetti 93/2, I-40129 Bologna, Italy\\
             \email{davide.massari@unibo.it} 
             \and
             INAF - Osservatorio di Astrofisica e Scienza dello Spazio di Bologna, Via Gobetti 93/3, I-40129 Bologna, Italy
             \and
             Kapteyn Astronomical Institute, University of Groningen, NL-9747 AD Groningen, Netherlands
             \and
             INAF - Osservatorio Astrofisico di Arcetri, Largo E. Fermi 5, 50127 Firenze, Italy
             \and
             ONERA, The French Aerospace Laboratory BP. 72, F-92322 Chatillon Cedex, France
             \and
             Aix Marseille Univ., CNRS, CNES LAM, 38 rue F. Joliot-Curie, 13388 Marseille, France
             \and
             European Southern Observatory (ESO), Alonso de C\'ordova 3107, Vitacura, Casilla 19001, Santiago, Chile  
             \and
             Univ. Grenoble Alpes, CNRS, IPAG, F-38000 Grenoble, France
             \and
             INAF-Osservatorio Astronomico di Roma, via Frascati 33, 0040 Monte Porzio Catone, Italy 
             \and
             European Southern Observatory, Karl-Schwarzschild-Str.2, 85748 Garching bei M\"{u}nchen, Germany
              }

   \date{Received 19/12/2019; accepted 22/1/2020}


  \abstract
  {Precise photometric and astrometric measurements on astronomical images require an accurate knowledge of the Point Spread Function (PSF).
  When the PSF cannot be modelled directly from the image, PSF-reconstruction techniques become the only viable solution. So far, however, their performance on real
  observations has rarely been quantified.}
  {In this Letter, we test the performance of a novel hybrid technique, called PRIME, on Adaptive Optics-assisted SPHERE/ZIMPOL observations of the Galactic globular cluster NGC~6121.}
  {PRIME couples PSF-reconstruction techniques, based on control-loop data and direct image fitting performed on the only bright point-like source available in the 
  field of view of the ZIMPOL exposures, with the aim of building the PSF model.}
  {By exploiting this model, the magnitudes and positions of the stars in the field can be measured with an unprecedented precision, which surpasses that obtained
  by more standard methods by at least a factor of four for on-axis stars and by up to a factor of two on fainter, off-axis stars.}
  {Our results demonstrate the power of PRIME in recovering precise magnitudes and positions when the information directly coming from astronomical images is limited to
  only a few point-like sources and, thus,  paving the way for a proper analysis of future Extremely Large Telescope observations of sparse stellar fields or individual extragalactic objects.}

   
   \maketitle
%

\section{Introduction}

Accurate knowledge of the Point Spread Function (PSF) is a necessary requirement for performing high-quality photometric and astrometric
analysis on astronomical imaging.
Standard methods used to model the PSF directly from the image typically exploit the wealth of information available in the observations of
dense stellar systems like globular clusters. In this kind of imaging, each star is the representation of the same PSF, but at a different pixel-phase.
By selecting isolated stars with a high signal-to-noise ratio (S/N) and combining their light profiles together, it is thus possible to recover a
good model for the actual shape of the PSF, which is then fit to all of the other sources in the image in order to measure their positions and magnitudes
(see e.g.~\citealt{stetson87, diolaiti00, anderson00}).

However, these standard methods are becoming less effective as we approach the era of Extremely Large Telescopes (ELTs), when observations will be routinely assisted by 
Adaptive Optics (AO) techniques. In fact, these techniques will enable diffraction-limited observations from the ground, at the cost of making the PSF variable across
the field of view (FoV). Because of the variability, fewer stars restricted to smaller portions of the FoV can be used as independent representations of the same PSF.
This is why, in sparse fields, very often it is the case that no stars at all other than the natural guide star will be available to model the PSF. 
This is when techniques for modelling the PSF which do not use the information coming from the images become most significant, such as PSF-reconstruction (PSF-R, \citealt{veran97}).

PSF-R is a technique that relies on AO control-loop data to determine the shape of the PSF potentially at any spatial location in the FoV. 
Despite its being theoretically well established (e.g. \citealt{joli18, ragland18}), so far, PSF-R has never surpassed the performance obtained 
by standard methods when applied to real astronomical imaging, not even in the case of AO-assisted data (e.g.~\citealt{turri15, massari16a, massari16b, monty18}).

In this Letter, we exploit a new hybrid technique known as PRIME (\citealt{beltramo19}), which combines PSF-R with image fitting to perform
a photometric and astrometric analysis of a small stellar field in the Galactic globular cluster NGC~6121 observed with SPHERE/ZIMPOL (\citealt{schmid18}).
While the theoretical background has been provided in Beltramo-Martin et al. (2020; hereafter BMS), here we focus on the results in terms of the
photometric and astrometric precision achieved on these real data. The comparison with results obtained using standard PSF-modelling methods shows, for the first
time, a significant improvement and demonstrates the potential of PRIME towards the advent of future ELT observations.

The Letter is organised as follows. In Section~\ref{data}, the observations of NGC~6121 that form the basis of this work are presented. In Section~\ref{onaxis}, we describe the methods
employed to perform the analysis and we show the results obtained on the on-axis guide star. In Section~\ref{off}, the photometric and astrometric precision is quantified for 
the off-axis sources. Finally, a discussion and the conclusions are provided in Section~\ref{concl}.

\section{Data analysis}\label{data}

In this work, we focus on imaging data of the Galactic globular cluster NGC~6121 taken with SPHERE/ZIMPOL as part of a technical proposal following the ESO Calibration Workshop 
2017\footnote{{\tt www.eso.org/sci/meetings/2017/calibration2017.html}}, under the program ID 60.A-9801(S).
The dataset consists of 12 exposures in the V filter, each with a duration of $200$ s (NDIT$=2$, DIT$=100$ s). Since ZIMPOL is capable of performing simultaneous observations through two cameras, we effectively end up with
24 independent measurements of the positions and magnitude of the stars in the FoV. Observations were taken on the night of 26 of June 2018, at an average airmass of $\sim1.2$.

\begin{figure}
    \includegraphics[width=\columnwidth]{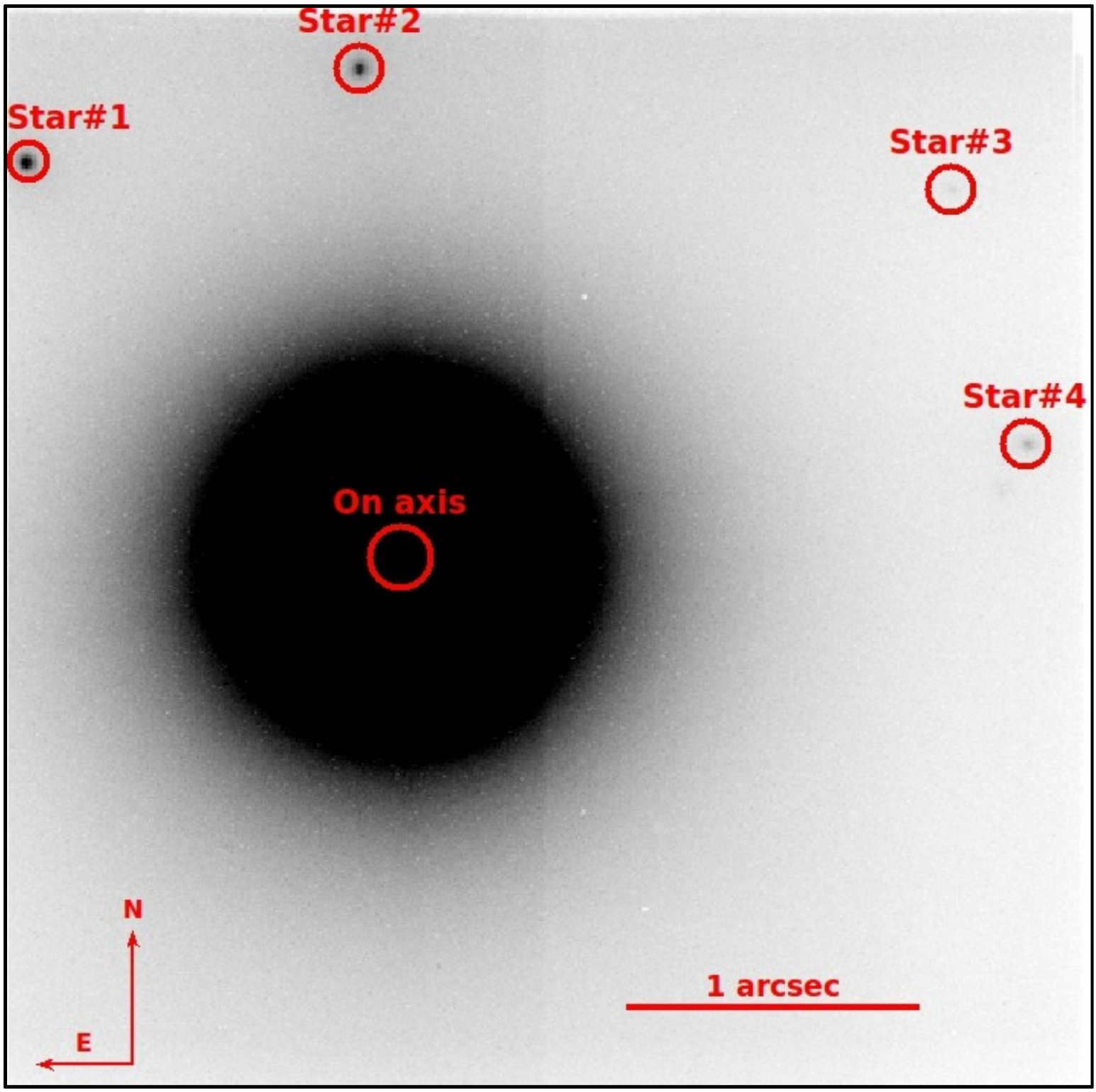}
    \caption{Median of the 24 exposures sampling the field of NGC~6121. The five detected stars are marked with red symbols. The orientation of the field and the physical scale 
    are also highlighted.}\label{fov}
\end{figure}

The FoV is a small squared window with a size of $3.5\times3.5$ arcsec$^2$, sampled with a pixel-scale of $7.2$ mas/pixel after data processing. Overall, five stars have been detected within the field. The brightest one
(V$\simeq10.9$ mag, \citealt{anderson08}) has been used as the natural guide star for the Adaptive Optics system, which has provided a fairly good correction throughout the observations, with an average Strehl ratio
in H-band of SR$=65$\% (this corresponding to SR$\sim2$\% in the adopted V-band) and an on-axis full-width at half maximum (FWHM) of FWHM$=33$ mas.
The other four stars are significantly fainter, having magnitudes in the range of V$=16.6-18.5$ and  with a position located at a distance of about 2 arcsec from the guide star.
For the sake of visualisation, in Fig.~\ref{fov} we show the averaged image of all of the available exposures, with the location of the five stars are marked in red. 

The data reduction, including bias subtraction and flat-fielding, was performed using the SPHERE Data and Reduction Handling
pipeline (\citealt{pavlov08}). Bad-pixel correction was applied using dedicated Python routines, as described in BMS.

The photometric and astrometric analysis is performed in two ways. The first uses standard PSF modelling techniques that only exploit the information
coming directly from the images and relies on the DAOPHOT-II suite of software (\citealt{stetson87}). The second uses the hybrid method encoded in PRIME instead.
In all of the exposures, DAOPHOT-II can only use the on-axis guide star to model the PSF as all of the other sources are too faint to actually provide further constraints
to the model. The way DAOPHOT-II works is described in detail in \cite{stetson87}. Briefly, the PSF is modelled by means of a Moffat function plus a table of residuals,
both of which are determined by fitting the stellar light profile enclosed within an aperture of an eight-pixel radius. Then the PSF that has been
recovered in this way is applied via ALLSTAR to all of the five stars within an aperture of a 30-pixel radius and, in our case, without any possibility for 
including PSF spatial variation in the measurements of magnitude and position. 
We note that this is a rather extreme case for resolved stellar population science that usually can rely on the presence of hundreds or thousands of stars in the field
to model the PSF. However, this is not uncommon at all in the investigations of individual extragalactic objects, such as the lensed Active Galactic Nuclei 
(e.g. \citealt{auger10, spingola19}), or in planetary science (e.g. \citealt{bonnefoy11}). This is why, despite the conditions are particularly poor to
perform the analysis with DAOPHOT-II, it is still important to compare its results with those coming from a better-suited method like PRIME.

On the other hand, PRIME exploits atmospheric and AO control-loop data to build a first-guess PSF model,
whose parameters are then adjusted based on the fitting of the natural guide star. Also in this case, the same model is then fit to all of the five stars.
Atmospheric parameters were simultaneously measured by the MASS-DIMM at Paranal (\citealt{butterley18, tokovinin07}), 
by the stereo-SCIDAR (\citealt{osborn18}), and the AO real time computer known as SPARTA (\citealt{suarezvalles12}), which all confirmed
that the observing conditions had been stable over the night, with the seeing at zenith varying between 0.7-0.9 arcsec.
Two 30 sec-long full sets of AO telemetry data were also obtained at the beginning of ZIMPOL observations (see BMS for details).

Finally, for the sake of comparison we also analysed the data on-axis using a simple aperture method. The magnitude measured in this way is given by the sum 
of the counts within a circular aperture of a 30-pixel radius in order to be consistent, thus, with the prescriptions used in the other two methods. 
Positions were derived within the same aperture radius as intensity-weighted centroids. All the individual exposures were treated independently during the data reduction and the analysis that followed. 

\section{Results: on-axis performance}\label{onaxis}

\begin{figure*}
    \includegraphics[width=\textwidth]{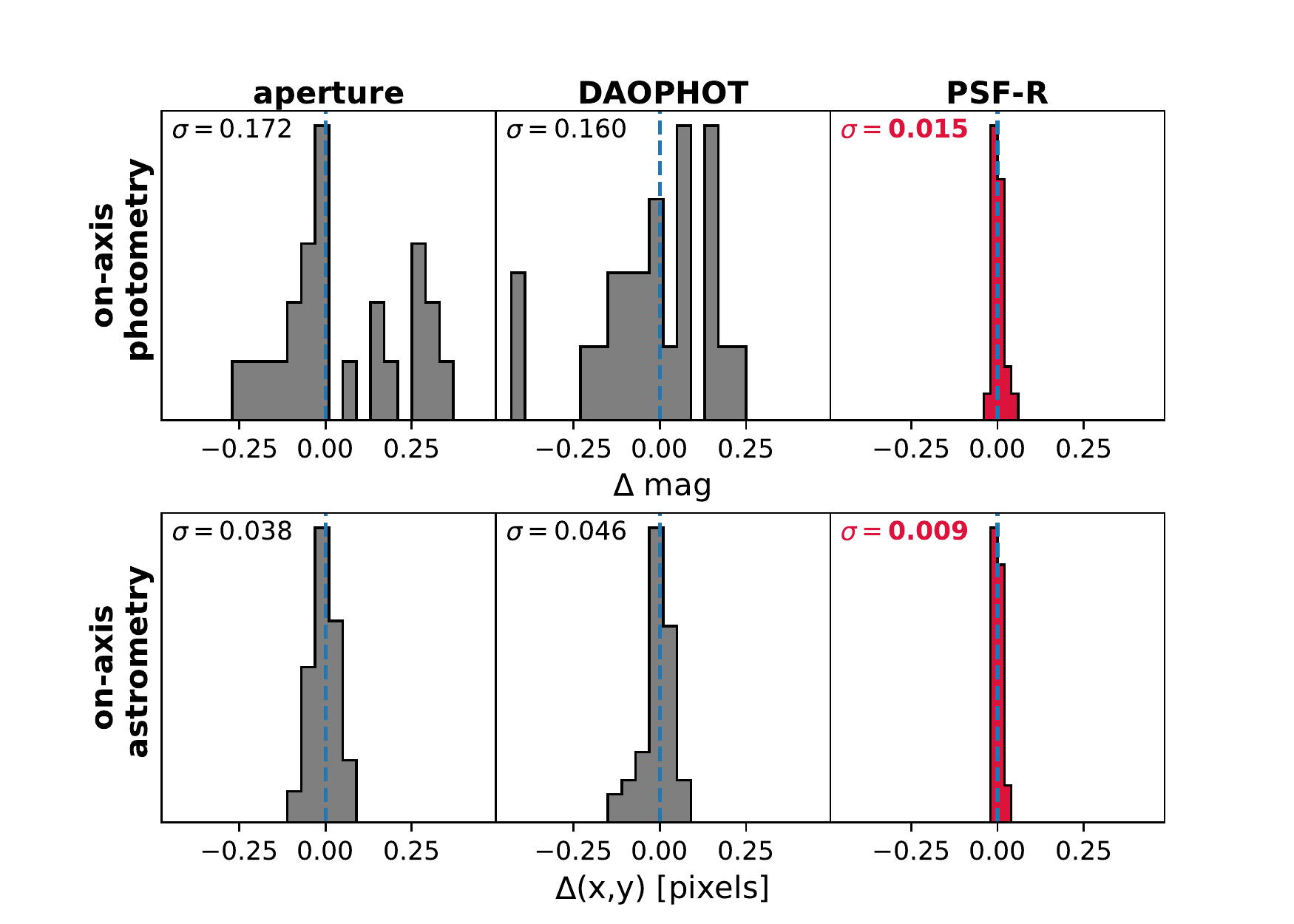}
    \caption{Photometric (top panel) and astrometric (bottom panel) precision achieved on the guide star when using the aperture method (left column), 
    DAOPHOT-II (middle column) and PRIME (right column).}\label{on}
\end{figure*}

The first target of the analysis is the natural guide star that has been used to model the PSF by DAOPHOT-II and to adjust the reconstructed PSF model by PRIME.
The results of the analysis are shown in Fig. \ref{on}, which compares the photometric (top panels) and astrometric (bottom panels) precision achieved
with the three methods employed. The precision ($\sigma$) is defined as the root mean square error around the mean value of the distributions of the $24$ independent magnitude and position
measurements of the guide star. We remark that a small degree of intrinsic photometric variation could be due to the temporal changes in the atmospheric conditions,
for example, because the airmass of our exposures ranges from $1.1$ at the beginning of the observations to $1.3$ at the end. However, this intrinsic variation
should be very small (sky transparency during the night was stable within a 2\% level and the airmass changed by very little) and, in any case, it affects the measurements 
obtained with the three methods in the same way. Moreover, the magnitudes measured in the two ZIMPOL cameras are slightly offset due to the different throughput of the two
channels. We corrected for such an offset by imposing that the average magnitudes of the guide star over the 12 exposures observed with each cameras are the same.

In terms of photometry, the aperture method and DAOPHOT-II perform similarly well, achieving a precision of $\sigma=0.172$ mag and $\sigma=0.160$ mag, respectively.
This may seem surprising as PSF-fitting usually produces much more precise results than aperture photometry. However, it should be considered that
the guide star is very bright and isolated, which are also the circumstances under which aperture photometry performs well.
On the other hand, it is interesting to note that PRIME achieves a precision one order of magnitude better than the previous two methods ($\sigma=0.015$ mag).
This level of precision matches the typical performance obtained on more stable, seeing-limited observations of dense stellar fields, where 
the modelling of the PSF is strongly facilitated by the large number of sources describing the same PSF (e.g. \citealt{massarim3, savino18}). 

In terms of astrometry, the precision achieved by the aperture method ($\sigma=0.038$ pixels) and DAOPHOT-II ($\sigma=0.046$ pixels) are again comparable for the same
reasons as those described in the photometry case. Moreover, also in this case, PRIME yields the best performance, achieving a precision of $\sigma=0.009$ pixels. The improvement,
therefore, amounts to about a factor of four compared to the other methods. When translated to an angular size, such a precision corresponds to $\sim70 ~\mu$as. 
This is a remarkable result as this level of performance is higher than that achieved so far by other Adaptive Optics facilities, which have otherwise achieved precision of the order of $150-200 ~\mu$as, at best, on resolved stellar population science cases 
 (see e.g. \citealt{ghez08, neichel14, massari16b}). Our findings match the results typically obtained by the best space-based 
astrometric facilities, such as {\it Gaia} (\citealt{brown18}) and the {\it Hubble Space Telescope} (e.g. \citealt{bellini14}).

\section{Results: off-axis performance}\label{off}

Four other stars, other than the guide star, are located in the FoV. They represent a very good opportunity to test PRIME in different conditions as they
are off-axis by about 2 arcsec and because they are much fainter than the guide star. Star-3 is in fact $\sim8$ mag fainter (see Fig.~\ref{fov}) and this is why 
in some of the exposures, its peak falls below the DAOPHOT-II detection limit. In these cases (17 out of 24 exposures), neither photometry nor astrometry for Star-3 were measured by DAOPHOT-II.
Moreover, since the aperture method is poorly suited to perform the analysis in the case of very faint stars, we restrict the following discussion 
to DAOPHOT-II and PRIME results only. In both cases, the positions and magnitude of the off-axis stars were measured after the subtraction of the on-axis star model.

\begin{figure}
    \includegraphics[width=\columnwidth]{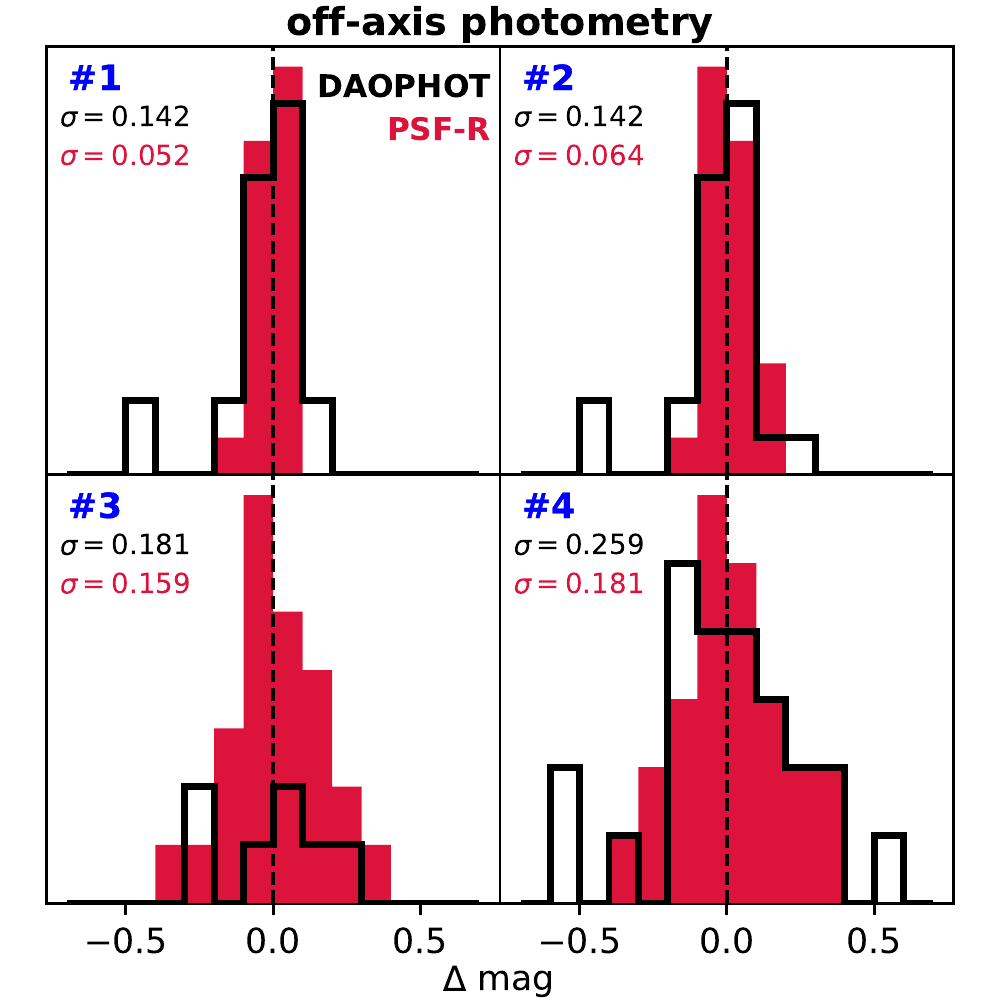}
    \caption{Comparison between the photometric precision achieved by DAOPHOT-II (black empty histograms) and PRIME (red histograms) on the four
    individual off-axis stars (each shown in a different panel). The values of the precision are also quoted for sake of comparison.}\label{offphoto}
\end{figure}

Figure~\ref{offphoto} shows the comparison between the photometric precision achieved by the two methods (black empty histograms for DAOPHOT-II measurements and 
red histograms for PRIME estimates) for each of the four off-axis stars.
Also in this case, PRIME seems to give better results compared to DAOPHOT-II, though not as strikingly as on the natural guide star. 
The photometric precision ranges from $\sigma\sim0.05$ mag for the two brightest sources (Star-1 and Star-2) up to $\sigma\sim0.17$ mag for the two faintest ones (Star-3 and Star-4).
The gain with respect to standard methods thus amounts to a factor of two to three in the best cases, while the performance is comparable for the faintest stars.
 
Existing Adaptive Optics photometry has been usually performed on K-band images. To compare it with our results, we first consider theoretical isochrones (\citealt{basti06})
for an old, metal-intermediate (\citealt{marino08}) globular cluster to find that the V-band magnitudes of our four off-axis stars lie in the range K$=15-16.5$ mag.
At these levels of K-band brightness and with similar exposure times ($\sim160$ s), the Multi-Conjugate Adaptive Optics (MCAO) camera GeMS/GSAOI (\citealt{rigaut12}) has
achieved a photometric precision of $\sim0.06$ mag (see \citealt{massari16a, saracino16}), which is remarkably similar to the PRIME performance described in this work. Nonetheless, we stress 
that our results were obtained in much more challenging conditions as $i$) we could only rely on one star to model the PSF (compared to the hundreds available in the quoted papers); $ii$) the SR of our V-band images is $\sim2$\% compared to SR$=20-30$\% for the GeMS IR data (e.g. \citealt{neichel14, massari16a, dalessandro16}); and $iii$) the PSF stability granted by 
GeMS MCAO system ($\sim15$\% FWHM variability over a $85\times85$ arcsec$^2$ FoV, see \citealt{bernard16}) is better than the one achieved by ZIMPOL ($\sim15$\% FWHM variability but over a much smaller
$3.5\times3.5$ arcsec$^2$ FoV).

\begin{figure}
    \includegraphics[width=\columnwidth]{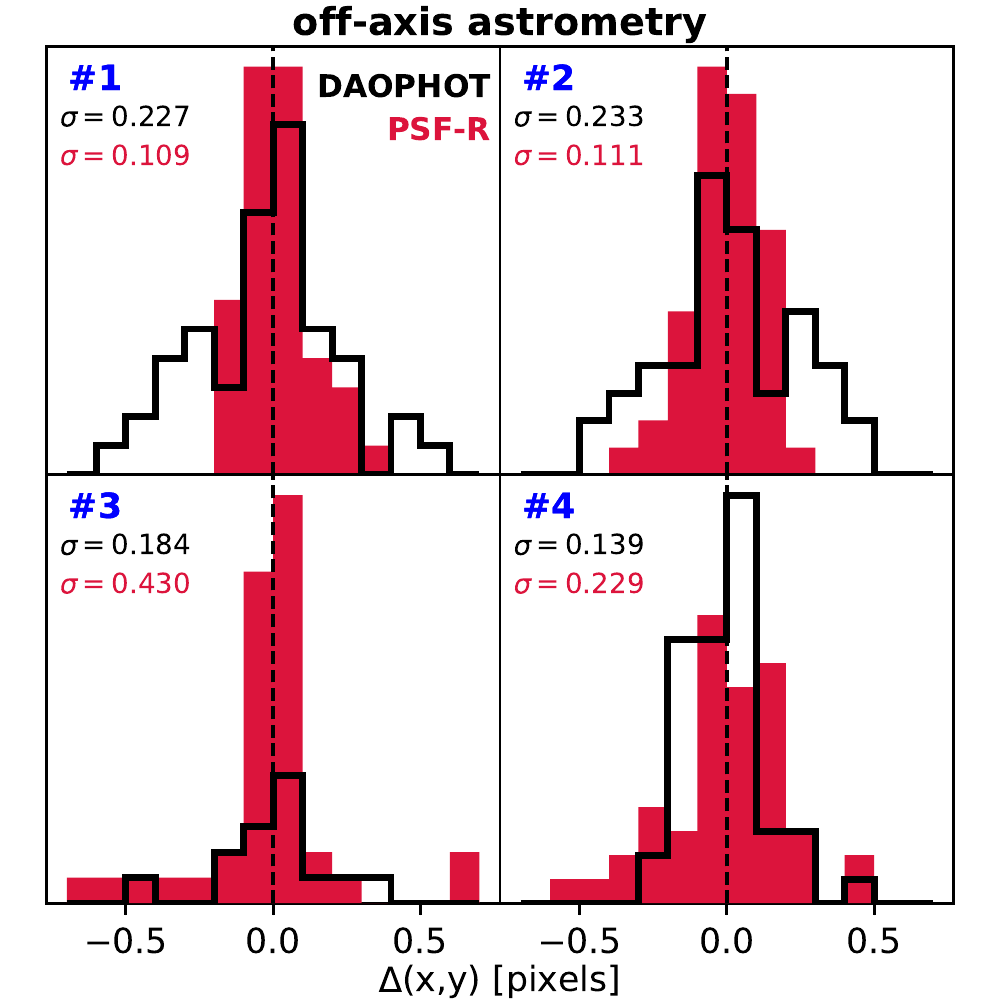}
    \caption{Astrometric precision achieved on off-axis stars with DAOPHOT-II and PRIME (the colour-coding is the same as in Fig.\ref{offphoto}).}\label{offastro}
\end{figure}

On the other hand, Fig.~\ref{offastro} compares the results obtained in terms of astrometric precision. As in the case of the guide star,
PRIME seems to maintain some advantage over DAOPHOT-II in recovering precise stellar positions. The improvement decreases, however, when moving from a factor
of four on-axis to a factor of $\sim2$ for the two brightest off-axis stars and to then disappear for the two faintest sources.
In the case of Star-3, the performance achieved by DAOPHOT-II seems to be a factor of two better. Although it is based on only a few measurements and, thus, could be affected by
small number statistics issues, it is, rather, the PRIME precision that is worse than expected. The large dispersion of PRIME measurements seems to be driven by few outliers, with the 
bulk of the distribution overlapping nicely with that from DAOPHOT-II. We therefore assess the two distributions as comparable.

Finally, also in this case we can compare our astrometric precision with that achieved by GeMS/GSAOI MCAO observations (having similar exposure times) on 
another Galactic globular cluster (NGC~6681, see \citealt{massari16b}). 
For stars in the same magnitude range, GeMS obtained a positional precision of the order of $0.5-1$ mas. Despite the lower S/N of our images caused by the less efficient 
AO correction in the optical bands, these results match ours well as the precision we achieved on Star-1 and Star-2 amounts to $\sim0.7$ mas.

\section{Summary and Conclusions}\label{concl}

In this Letter, we carried out an experimental comparison of the photometric and astrometric performance achieved using both standard PSF-modelling techniques and the PSF-R-based hybrid technique known as PRIME 
on real observations of a small field in the Galactic globular cluster NGC~6121.
The conditions of the experiment are rather extreme, with only one star in the field bright enough to be suitable for PSF-modelling. Despite
being a rather uncommon condition in the case of resolved stellar populations studies, such an occurrence is more typical for investigations of individual extragalactic objects
and it is, therefore, valuable for a wealth of science cases.

On-axis, the hybrid approach followed by PRIME that starts from a PSF model reconstructed using AO control-loop data and then adjusts its parameters
via image fitting (see the detailed description of the method in \citealt{beltramo19}) achieves remarkable results. The photometric precision
is $\sigma=0.015$ mag, about one order of magnitude better than what is obtained using more standard methods like aperture photometry or DAOPHOT-II.
Also in terms of astrometric precision, the performance ensured by PRIME is excellent ($\sigma\simeq70~\mu$as), leading to a gain of about a factor of four
compared to the other methods.

On the other hand, when assessing the performance on the four, faint off-axis sources, PRIME achieves results that are comparable with those obtained
by DAOPHOT-II on both photometry and astrometry. When the comparison refers to existing analysis of GeMS MCAO observations of stars in a similar magnitude range, 
the results in terms of precision are again similar, despite the unfavourable conditions to perform the analysis. 

In conclusion, these results demonstrate the potential of PRIME as a powerful tool for analysing future Adaptive Optics-assisted observations with  ELTs. This is especially true 
in cases where there is a poor availability of point-like sources in the field for modelling the PSF. Future experiments will test PRIME performance on observational cases that
are well suited for an analysis with standard methods as well. 

\begin{acknowledgements}

Based on observations collected at the European Organisation for Astronomical Research in the Southern Hemisphere.
GF has been supported by the Futuro in Ricerca 2013 (grant RBFR13J716).

\end{acknowledgements}


\begin{thebibliography}{}

\bibitem[Alonso et al.(1999)]{1999A&AS..140..261A} Alonso, A., Arribas, S., \& Mart{\'\i}nez-Roger, C.\ 1999, \aaps, 140, 261

\bibitem[Anderson, \& King(2000)]{anderson00} Anderson, J., \& King, I.~R.\ 2000, \pasp, 112, 1360

\bibitem[Anderson et al.(2008)]{anderson08} Anderson, J., Sarajedini, A., Bedin, L.~R., et al.\ 2008, \aj, 135, 2055

\bibitem[Auger et al.(2010)]{auger10} Auger, M.~W., Treu, T., Bolton, A.~S., et al.\ 2010, \apj, 724, 511

\bibitem[Bellini et al.(2014)]{bellini14} Bellini, A., Anderson, J., van der Marel, R.~P., et al.\ 2014, \apj, 797, 115

\bibitem[Beltramo-Martin et al.(2019)]{beltramo19} Beltramo-Martin, O., Correia, C.~M., Ragland, S., et al.\ 2019, \mnras, 487, 5450

\bibitem[Beltramo-Martin et al.(2020)]{bms} Beltramo-Martin, O., Marasco, A., Fusco, T., et al.\ 2020, \mnras, submitted

\bibitem[Bernard et al.(2016)]{bernard16} Bernard, A., Neichel, B., Samal, M.~R., et al.\ 2016, \aap, 592, A77

\bibitem[Bonnefoy et al.(2011)]{bonnefoy11} Bonnefoy, M., Lagrange, A.-M., Boccaletti, A., et al.\ 2011, \aap, 528, L15

\bibitem[Butterley et al.(2018)]{butterley18} Butterley, T., Sarazin, M., Navarrete, J., et al.\ 2018, \procspie, 107036G

\bibitem[Dalessandro et al.(2016)]{dalessandro16} Dalessandro, E., Saracino, S., Origlia, L., et al.\ 2016, \apj, 833, 111

\bibitem[Diolaiti et al.(2000)]{diolaiti00} Diolaiti, E., Bendinelli, O., Bonaccini, D., et al.\ 2000, \aaps, 147, 335

\bibitem[Gaia Collaboration et al.(2018)]{brown18} Gaia Collaboration, Brown, A.~G.~A., Vallenari, A., et al.\ 2018, \aap, 616, A1

\bibitem[Ghez et al.(2008)]{ghez08} Ghez, A.~M., Salim, S., Weinberg, N.~N., et al.\ 2008, \apj, 689, 1044

\bibitem[Jolissaint et al.(2018)]{joli18} Jolissaint, L., Ragland, S., Christou, J., et al.\ 2018, \ao, 57, 7837

\bibitem[Marino et al.(2008)]{marino08} Marino, A.~F., Villanova, S., Piotto, G., et al.\ 2008, \aap, 490, 625

\bibitem[Massari et al.(2016a)]{massari16a} Massari, D., Fiorentino, G., McConnachie, A., et al.\ 2016, \aap, 586, A51

\bibitem[Massari et al.(2016b)]{massari16b} Massari, D., Fiorentino, G., McConnachie, A., et al.\ 2016, \aap, 595, L2

\bibitem[Massari et al.(2016c)]{massarim3} Massari, D., Lapenna, E., Bragaglia, A., et al.\ 2016, \mnras, 458, 4162

\bibitem[Monty et al.(2018)]{monty18} Monty, S., Puzia, T.~H., Miller, B.~W., et al.\ 2018, \apj, 865, 160

\bibitem[Neichel et al.(2014)]{neichel14} Neichel, B., Lu, J.~R., Rigaut, F., et al.\ 2014, \mnras, 445, 500

\bibitem[Pavlov et al.(2008)]{pavlov08} Pavlov, A., M{\"o}ller-Nilsson, O., Feldt, M., et al.\ 2008, \procspie, 701939

\bibitem[Osborn et al.(2018)]{osborn18} Osborn, J., Wilson, R.~W., Sarazin, M., et al.\ 2018, \mnras, 478, 825

\bibitem[Pietrinferni et al.(2006)]{basti06} Pietrinferni, A., Cassisi, S., Salaris, M., et al.\ 2006, \apj, 642, 797

\bibitem[Ragland et al.(2018)]{ragland18} Ragland, S., Dupuy, T.~J., Jolissaint, L., et al.\ 2018, \procspie, 107031J

\bibitem[Rigaut et al.(2012)]{rigaut12} Rigaut, F., Neichel, B., Boccas, M., et al.\ 2012, \procspie, 84470I

\bibitem[Saracino et al.(2016)]{saracino16} Saracino, S., Dalessandro, E., Ferraro, F.~R., et al.\ 2016, \apj, 832, 48

\bibitem[Sarajedini et al.(2007)]{sarajedini07} Sarajedini, A., Bedin, L.~R., Chaboyer, B., et al.\ 2007, \aj, 133, 1658

\bibitem[Savino et al.(2018)]{savino18} Savino, A., Massari, D., Bragaglia, A., et al.\ 2018, \mnras, 474, 4438

\bibitem[Schmid et al.(2018)]{schmid18} Schmid, H.~M., Bazzon, A., Roelfsema, R., et al.\ 2018, \aap, 619, A9

\bibitem[Spingola et al.(2019)]{spingola19} Spingola, C., McKean, J.~P., Massari, D., et al.\ 2019, \aap, 630, A108

\bibitem[Stetson(1987)]{stetson87} Stetson, P.~B.\ 1987, \pasp, 99, 191

\bibitem[Su{\'a}rez Valles et al.(2012)]{suarezvalles12} Su{\'a}rez Valles, M., Fedrigo, E., Donaldson, R.~H., et al.\ 2012, \procspie, 84472Q

\bibitem[Tokovinin, \& Kornilov(2007)]{tokovinin07} Tokovinin, A., \& Kornilov, V.\ 2007, \mnras, 381, 1179

\bibitem[Turri et al.(2015)]{turri15} Turri, P., McConnachie, A.~W., Stetson, P.~B., et al.\ 2015, \apjl, 811, L15

\bibitem[Veran et al.(1997)]{veran97} Veran, J.-P., Rigaut, F., Maitre, H., et al.\ 1997, Journal of the Optical Society of America A, 14, 3057

\end{thebibliography}
\end{document}